\documentclass[authoryear,review,11pt]{elsarticle}
\usepackage{amsmath,amsfonts}

\newcommand{\ud}{\mathrm{d}}

\journal{\tiny Journal of the Royal Society Interface}
\begin{document}

\begin{frontmatter}
  \title{Quantifying Limits to Detection of Early Warning for Critical Transitions}
  \author[cpb]{Carl Boettiger\corref{cor1}}
  \ead{cboettig@ucdavis.edu}
  \author[esp]{Alan Hastings}
  \cortext[cor1]{Corresponding author.}
  \address[cpb]{Center for Population Biology, 1 Shields Avenue, University of California, Davis, CA, 95616 United States.}
  \address[esp]{Department of Environmental Science and Policy, University of California, Davis, CA, 95616 United States}

  \begin{abstract}

Catastrophic regime shifts in complex natural systems may be averted through advanced detection.
Recent work has provided a proof-of-principle that many systems approaching a catastrophic transition may be identified
through the lens of early warning indicators such as rising variance or increased return times.
Despite widespread appreciation of the difficulties and uncertainty involved in such forecasts,
proposed methods hardly ever characterize their expected error rates. 
Without the benefits of replicates, controls, or hindsight, 
applications of these approaches must quantify how reliable different indicators are in avoiding false alarms,
and how sensitive they are to missing subtle warning signs.
We propose a model based approach in order to quantify this trade-off between reliability and sensitivity
and allow comparisons between different indicators.
We show these error rates can be quite severe for common indicators even under favorable assumptions,
and also illustrate how a model-based indicator can improve this performance. 
We demonstrate how the performance of an early warning indicator varies in different data sets, 
and suggest that uncertainty quantification become a more central part of early warning predictions.  
  \end{abstract}

  \begin{keyword}
early warning signals \sep tipping point \sep alternative stable states \sep likelihood methods
   \end{keyword}
 \end{frontmatter}

\section{Introduction}
There is an increasing recognition of the importance of regime shifts or critical transitions at a variety of scales in ecological systems~\citep{Holling1973, Wissel1984, Scheffer2001, Scheffer2009, Drake2010, Carpenter2011}⁠.
Many important ecosystems may currently be threatened with collapse, including corals~\citep{Bellwood2004}, fisheries~\citep{Berkes2006}⁠, lakes~\citep{Carpenter2011}, and semi-arid ecosystems~\citep{Kefi2007}⁠.
Given the potential impact of  these shifts on the sustainable delivery of ecosystem services~\citep{Folke2004}
and the need for management to either avoid an undesirable shift or else to adapt to novel conditions,
it is important to develop the ability to predict impending regime shifts based on early warning signs.

A number of particular systems have demonstrated the kinds of relationships that would produce regime shifts,
including dynamics of coral reefs~\citep{Mumby2007},
and simple models of metapopulations with differing local population sizes~\citep{Hastings1991a}.
In cases like these one sensible approach to understanding 
whether a regime shift would be likely would be 
to fit the model using either a time series or else independent estimates of parameters. 
More generally, with a good model of the system, detail-oriented approaches could be useful~\citep{Lade2012}.  
In this treatment we focus on the situation where these more detailed models are not available.  

Indeed, for many ecological systems specific models are not available and general approaches are needed~\citep{Scheffer2009, Lade2012}
that do not depend on estimating the parameters of a known model of a specific system. 
This has led to a variety of approaches based on summary statistics
~\citep[\emph{e.g.}][]{Carpenter2006, Held2004, Dakos2008, Guttal2008, Biggs2009, Carpenter2011, Seekell2011}
that look for generic signs of impending regime shifts.  
Here we extend earlier work by providing estimates 
of the ability of different potential indicators to accurately signal impending regime shifts,
and develop new approaches that both are more efficient
and also lay bare some of the important assumptions underlying attempts 
to find general warning signs of regime shifts.  
We distinguish this question from the extensive literature involving change-point analysis 
for the post-hoc identification of if and when a regime shift has occurred~\citep{Easterling1995, Rodionov2004, Lenton2009}, 
which is of little use if the goal is the advanced detection of the shift.

We begin by discussing the limitations of current approaches that rely on summary statistics
and provide a description of assumptions through the introduction of a model based approach to detect early warning signals.
We then illustrate how stochastic differential equation (SDE) models can be used
to reflect the uncertainty inherent in the detection of early warning signals. 
We caution against paradigms that are not useful for capturing uncertainty in
a model-selection based approach, such as information criteria.
Finally we use receiver-operating characteristics~\citep{Green1989, Keller2009}
as a way to illustrate the sensitivity different data sets
and different indicators have in detecting early warning signals and use this to explore a number of examples.
This approach provides a visualization of the types of errors that arise and how one can trade off between them,
and is important for framing the problem as one focused on prediction.

\section{The summary statistics approach}
Foundational work on early warning signals has operated under the often-implicit assumption
that the system dynamics contain a saddle-node bifurcation by
looking for patterns that are associated with this kind of transition.
A saddle-node bifurcation occurs when a parameter changes and a stable equilibrium 
(node) and an unstable equilibrium (saddle) coalesce and dissapear.  The system
then moves to a more distant equilbrium.  \citet{Guckenheimer1983} or any other 
textbook on dynamical systems will provide precise definitions and further explanation.   

Typical patterns used as warning signals include an increasing trend in a summary statistic such as
variance~\citep{Carpenter2006}, autocorrelation~\citep{Held2004, Dakos2008},
skew~\citep{Guttal2008}, spectral ratio~\citep{Biggs2009}.
While attractive for their simplicity, such approaches must confront numerous challenges.
In this paper we argue for a model-based approach to warning signals,
and describe how this can be done in a way that best addresses these difficulties.
We begin by enumerating several of the difficulties encountered in approaches lacking an explicit model.

\subsubsection*{Hidden assumptions}
The underlying assumption that the system contains a saddle-node bifurcation
can be easily overlooked in common summary-statistics based approaches.
For instance, variance may increase for reasons that
do not signal an approaching transition~\citep{Schreiber2003, Schreiber2008}.
Alternatively, variance may not increase as a bifurcation is approached~\citep{Livina2012, Dakos2011a}.
Some classes of sudden transitions may exhibit no warning signals~\cite{Hastings2010}.
Like saddle-node bifurcations, transcritical bifurcations involve an eigenvalue passing through zero,
and exhibit the patterns of critical slowing down and increased variance~\citep{Drake2010}.  
However, transcritical bifurcations involve a change in stability of a fixed point, rather
than the sudden dissapearance of a fixed point that has made critical transitions so worrisome.
While no approach will be applicable to all classes of sudden transitions,
it is certainly still useful to have an approach that detects transitions driven by
saddle-node bifurcations, which have been found in many contexts~\citep[\emph{e.g.}, see][]{Scheffer2001}.

Even when we can exclude or ignore other dynamics and
restrict ourselves to systems that can produce a saddle-node bifurcation,
approaches based on critical slowing down or rising variance~\citep[\emph{e.g.}][]{Held2004, Scheffer2009, Carpenter2011}
must further assume that a changing parameter has brought the system closer to the bifurcation.
This assumption excludes at least three alternative explanations for the transition in system behavior.
The first possibility is that a large perturbation of the system state
has moved the system into the alternative basin of attraction~\citep{Scheffer2001}.
This is an exogenous forcing that does not arise from the system dynamics, so it is not the kind of event we can expect to forecast.
(An example might be a sudden dramatic increase in fishing effort that pushes a harvested population past a threshold.)
The second scenario is a purely noise-induced transition, a chance fluctuation that happens to carry the system across the boundary~\citep{Ditlevsen2010}.
\citet{Livina2012} indicate that such noise induced transitions cannot be predicted through early warning signals -- 
at least they are not expected to exhibit the same early warning patterns of increased variance and increased autocorrelation
anticipated in the case of a saddle-node bifurcation.
The third scenario is that the system does pass through a saddle-node bifurcation,
but rather than gradually and monotonically approaching the critical point, the
bifurcation parameter moves in a rapid or highly non-linear way, making the detection of any gradual trend impossible.



\subsubsection*{Arbitrary windows}
In addition to the assumption of a saddle-node bifurcation, 
the calculation of statistics that would be used to detect an impending transition is subject to several arbitrary choices.
A basic difficulty arises from the need to assume a time-series is \emph{ergodic}: 
that averaging over time is equivalent to averaging over replicate realizations,
while trying to test if it is not.
Theoretically, the increasing trend in variance, autocorrelation, or other statistics
is something that would be measured across an ensemble -- across replicates. 
As true replicates are seldom available in systems for which developing warning signals would be most desirable,
typical methods average across a single replicate using a moving window in time.
The selection of the size of this window and whether and by how much to overlap consecutive windows
varies across the literature.
\citet{Lenton2012} demonstrates that these differences can influence the results,
and that the different choices each carry advantages and disadvantages.

In addition to introducing the challenge of selecting a window size,
this ergodic assumption raises further difficulties.
While appropriate for a system that is stationary,
or changing slowly enough in the window that it may appear stationary,
the assumption is at odds with the original hypothesis
that the system is approaching a saddle-node bifurcation.

Further, certain statistics such as the critical slowing down measured by autocorrelation
require data that is evenly sampled in time.
Interpolating from existing data to create evenly spaced points is particularly problematic,
as this introduces an artificial autocorrelation into the data.

\subsubsection*{No quantitative measures}
Summary statistics typically invoke qualitative patterns such as an increase in statistic $x$,
rather than a quantitative measure of the early warning pattern.
This makes it difficult to compare between signals or to
attribute a statistical significance to the detection.
Some authors have suggested Kendall's correlation coefficient,
$\tau$, could be used to quantify an increase~\citep{Dakos2008, Dakos2011}
in autocorrelation or variance.
Other measures of increase, such as Pearson's correlation coefficient have also been proposed~\citep{Drake2010},
while most of the literature simply forgoes quantifying the increase or estimating significance.
While adequate in experimental systems that can compare patterns between controls 
and replicates~\citep[\emph{e.g.}][]{Drake2010, Carpenter2011}, 
any real-world application of these approaches must be useful on a single time-series of observations.
In these cases a quantitative definition of a statistically significant detection is essential.
Without this, we have no assurance that a purported detection is not, in fact, a false positive. 
By focusing primarily on examples known to be approaching a transition when testing warning signals,
the probability of false positives has largely been overlooked.  

\subsubsection*{Problematic null models}
Specifying an appropriate null model is also difficult.
Non-parametric null hypotheses seem to require the fewest assumptions but in fact can be the most problematic.
For instance, the standard non-parametric hypothesis test with Kendall's tau rank correlation coefficient
assumes only that the two variables are independent,
but this is an assumption that is violated by the very experimental design:
temporal correlations will exist in any finely-enough sampled time series,
and moving windows introduce temporal correlations in the statistics.
Under such a test any adequately large data set will find a significant result,
regardless of whether a warning signal exists.
A similar problem arises when the points in the time series are reordered to create a null hypothesis --
this destroys the natural autocorrelation in the time series.
More promising parametric null models have been proposed,
such as autoregressive models in~\citet{Dakos2008}, bringing us
closer to a model-based approach with explicit assumptions.
Others have looked for alternative summary statistics where
reasonable null models are more readily available,
such as~\citet{Seekell2011}'s proposal to test for conditional heteroscedasticity.

\subsubsection*{Summary-statistic approaches have less statistical power.}
Methods for the detection of early warning signals are continually challenged by inadequate data~\citep{Inman2011, Scheffer2010,Held2004, Dakos2008, Scheffer2009, Guttal2008, Carpenter2011, Bestelmeyer2011}.
Despite the widespread recognition of the this need for large data sets,
there has been very few studies quantitative studies of power to determine at how much data is required~\citep{Contamin2009},
how often a particular method would produce a false alarm or fail to detect a signal,
and which tests will be the most powerful or sensitive.
The Neyman-Pearson Lemma demonstrates that the most powerful test between hypotheses
compares the likelihood that the data was produced under each~\citep{Neyman1933}.
Such likelihood calculations require a model-based approach.

\section{A model based approach}
Model-based approaches are beginning to play a larger role in early warning signal detection,
though we have not as yet seen the direct fitting and simulation of models to compare hypotheses.
While choosing appropriate models without system-specific knowledge is challenging,
much can be accomplished by framing the implicit assumptions into equations.
\citet{Lade2012} introduce the idea of generalized models for early warning signals, and
\citet{Kuehn2011} presents normal forms for bifurcation processes that can give rise to critical transitions.
\citet{Carpenter2011e} and \citet{Dakos2011a} start by assuming the dynamics obey a generic
stochastic differential equation (SDE), but use this only to derive or define the summary statistics of interest.

In this section we outline how the detection of early warning signals may be thought of
as a problem of model choice.
We next show generic models can be constructed under the assumptions discussed above
and estimated from the data in a maximum likelihood framework.
We highlight the disadvantages of comparing these estimates by information criteria,
and instead introduce a simulation or bootstrapping approach rooted in~\citet{Cox1961} and~\citet{McLachlan1987} that
characterizes the rate of missed detections and false alarms expected in the estimate.

\subsection*{Early warning signals as model choice}
It may be useful to think of the detection of early warning signals
as a problem of model choice rather than one of pattern recognition.
The model choice approach attempts to frame each of the possible scenarios as structurally different equations,
each with unknown parameters that must be estimated from the data.
In any model choice problem, it is important to identify the goal of the exercise --
such as the ability to generalize, to imitate reality, or to predict~\citep{Levins1966}.
In this case generality is more important than realism or predictive capability:
we will write down a general model that is capable of approximating
a wide class of models in which regime shifts are characterized by a saddle-node bifurcation,
and a second generic model that is capable of representing the behavior of such systems
when they are not approaching a bifurcation.
These may be thought of as the hypothesis and null hypothesis,
though they are in fact compound hypotheses,
as we must first estimate the model parameters from the data.
In this approach it is not assumed that ``reality'' is included in the models being tested,
but that one of the models is a better approximation of the true dynamics than the other.
System whose dynamics violate the assumptions common to both models,
such as in the examples of ~\citet{Hastings2010} where systems exhibit sudden transitions without warning,
fall outside the set of cases where this approach would be valid;
though the inability of either model to match the system dynamics could be an indication of such a violation.

\subsection*{Models}
In the neighborhood of a bifurcation a system can be transformed into its \emph{normal form}
by a change of variables to facilitate analysis ~\citep{Guckenheimer1983}.
The normal form~\citep{Guckenheimer1983, Kuehn2011} for the saddle-node bifurcation is
\begin{equation}
\frac{\ud x}{\ud t} = r_t- x^2.
\label{saddle-node}
\end{equation}
where $x$ is the state variable and $r_t$ our bifurcation parameter.
We have added a subscript $t$ to the bifurcation parameter as a reminder that
it is the value which may be slowly varying in time and
consequently moving the system closer to a critical transition or regime shift~\citep{Scheffer2009}.
Transforming this canonical form to allow for an arbitrary mean in the state variable $\theta$,
the system near the bifurcation looks like \( dx/dt = r_t- (\theta-x)^2 \), with fixed point \(\hat x = \sqrt{r_t} +\theta =: \phi(r_t)\).
We expand around the fixed point and express as a stochastic differential equation~\citep[\emph{e.g.}][]{Gardiner2009}:

\begin{equation}
\ud X = \sqrt{ r_t } (\phi(r_t) - X_t)\ud t + \sigma\sqrt{\phi(r_t) } \ud B_t \label{LSN}
\end{equation}

where $B_t$ is the standard Brownian motion. 
This expression captures the behavior of the system near the stable point as it approaches the bifurcation.
Allowing the stochastic term to scale with the square root of $\phi$ follows from the assumption that
of an internal-noise process, such as demographic stochasticity, that arises in deriving the SDE from
a Markov process, see~\citet{Kampen2007a} or~\citet{Black2012}. 
The square root could be removed for an external noise process, such as environmental noise.
In practice it will be difficult to descriminate between the square root and linear scaling
in these applications, since the average value of the state changes little before the bifurcation.  


As we discuss above, in this paradigm we must include an assumption on how the bifurcation parameter, $r_t$, is changing.
We assume a gradual, monotonic change which we approximate to first order:

\begin{equation}
r_t = r_0 - m t.
\label{R_t}
\end{equation}

Detecting accelerating or otherwise nonlinear approaches to the bifurcation will generally require more power.
When the underlying system is not changing, $r_t$ is constant ($m=0$) and Equation~\eqref{LSN} will reduce to a simple Ornstein-Uhlenbeck process,

\begin{equation}
\ud X_t = r (\theta - X_t) \ud t + \sigma \ud B_t \label{OU}
\end{equation}

This is the continuous time analog of the first-order autoregressive model considered as a null model elsewhere~\citep[\emph{e.g.}][]{Dakos2008, Guttal2008a}.

\subsection*{Likelihood calculations}
The probability \(P(X|M) \) of the data \(X\) given the model \(M\) is the product of the probability of observing each point in the time series given the previous point and the length of the interval,

\begin{equation}
\log P(X | M)=  \sum_i \log P(x_i | x_{i-1}, t_i)
\end{equation}

For~\eqref{LSN} or~\eqref{OU} it is sufficient~\citep{Gardiner2009} to solve the moment equations for mean and variance respectively:

\begin{align}
 \frac{\ud }{\ud t} E(x| M)&=  f(x) \\
\frac{\ud}{\ud t} V(x| M) &=  -\partial_x f(x) V(x|M) + g(x)^2
  \label{general_moments}
\end{align}

For the OU process, we can solve this in closed form over an interval of time $t_i$ between subsequent observations:

\begin{align}
  E(x_i| M = \text{OU}) &= X_{i-1} e^{-r t_i} + \theta \left(1 - e^{-rt_i} \right) \\
V(x_i| M = \text{OU}) &= \frac{\sigma^2}{2 r} \left(1 - e^{-2 r t_i} \right)
\label{OUsoln}
\end{align}

For the time dependent model, we have analytic forms only for the dynamical equations of these moments from equation~\eqref{general_moments}, which we must integrate numerically over each time interval.
The moments of Equation~\eqref{LSN} are given by

\begin{align}
\frac{\ud }{\ud t} E(x_i| M = \text{LSN})&=  2\sqrt{r(t)}(\sqrt{r(t)}+\theta - x_i) \\
\frac{\ud}{\ud t} V(x_i| M = \text{LSN}) &=  -2 \sqrt{r(t)} V(x_i) + \sigma^2 ( \sqrt{r(t)}+\theta )
\label{LSNsoln}
\end{align}

These are numerically integrated using \texttt{lsoda} routine available in \texttt{R} for the likelihood calculation.

\subsection*{Comparing Models}
Likelihood methods form the basis of much of modern statistics
in both Frequentist and Bayesian paradigms.
The ability to evaluate likelihoods directly by computation has made it
possible to treat cases that do not conform to traditional assumptions more directly.
The basis of likelihood comparisons has its roots in the Neyman-Pearson Lemma,
which essentially asserts that comparing likelihoods is the most powerful test
of a choice between two hypotheses~\citep{Neyman1933}, and motivates
tests from the simple likelihood ratio test up through modern model adequacy methods.

The hypotheses considered here are more challenging then the original lemma provides for,
as they are composite in nature:
they specify two model forms (stable and changing stability)
but with model parameters that must be first estimated from the data.
Comparing models whose parameters have been estimated by maximum likelihood is first treated by~\citet{Cox1961, Cox1962},
and has since been developed in this simulation estimation of the null distribution~\citep{McLachlan1987}, by parametric bootstrap estimate~\citep{Efron1987}.
Cox's $\delta$ statistic (often called the deviance between models)
is simply the difference between the log likelihoods of these maximum likelihood estimates, defined as follows.

Let $L_0$ be the likelihood function for model 0,
let $\theta_0 = \arg \max \theta_0 \in \Omega_0$, ($L_0 (\theta_0 |X)$)
be the maximum likelihood estimator for $\theta_0$ given $X$, and let $L_0 = L_0 (\theta_0 |X)$;
and define $L_1$, $\theta_1$, $L_1$ similarly for model 1.
The statistic we will use is $\delta$,
defined to be twice the difference in log likelihood of observing the data under the two MLE models,
\begin{equation}\label{delta}
\delta = -2 (\log L_0 - \log L_1 ).
\end{equation}
This approach has been applied to the problem of model adequacy~\citep{Goldman1993}
and model choice~\citep{Huelsenbeck1996} in other contexts.
We have extended the approach by generating the test distribution
as well as a null distribution of the statistic $\delta$.

\subsection{Simulation-based comparisons} \label{simbased}
We perform the identical analysis procedure described above on each of these three data sets.
First, we estimate parameters for the null and test model to each data set by maximum likelihood.
Comparing the likelihood of these fits directly gives us only a minimal indication of which model fits better. 
To identify if these differences are significant,
and by what probability they could arise as a false alarm or a missed event,
we simulate 500 replicate time series from each estimated model.

The model parameters of both models are re-estimated on both families of replicates
(the null and test, \emph{i.e.} $2 \times 2 \times 500$ fits).
The differences in the likelihood values between the model estimates produced from the first set of simulations
determines the null distribution for the deviance statistic $\delta$.
As the constant OU process model is nested within the time-heterogeneous model, these values are always positive,
but tend to be not as large as those produced when the models are fit to the second family of data.

The extent to which these distributions overlap indicates our inability to distinguish between these scenarios.
The tendency of the observed deviance to fall more clearly in the domain of one distribution or the other
indicates the probability our observed data corresponds best with that model  --
either approaching a critical transition or remaining stable.
While it trivial to assign a statistical significance to this observation based
on how far into the tail of the null distribution it falls,
for the reasons we discussed we prefer the more symmetric comparison of the probability that
this value was observed in either distribution.
We visualize the trade-off between false alarms and failed detection using the ROC curves introduced above.

\subsection*{Information criteria will not serve.}
One will commonly observe models representing alternative processes being compared through
the use of various information criteria such as the Akaike information criterion.
While tempting to apply in this situation, such approaches are not suited to this problem for several reasons.
The first is that information criteria are not concerned with the model choice objective we have in mind,
as they are typically applied to find an adequate model description without too many parameters that the system may be over-fit.
More pointedly, information criteria have no inherent notion of uncertainty.
Information criteria tests alone will not tell us our chances of a false alarm, of missing a real signal,
or how much data we need to be confident in our ability to detect transitions.

\subsection*{Beyond hypothesis testing}
It is possible to frame the question of sensitivity, reliability, and adequate data in the language of hypothesis testing.
This introduces the need for selecting a statistical significance criterion.
In the hypothesis testing framework, a false positive is a Type I error,
which is defined relative to this arbitrary statistical significance criterion, most commonly 0.05.
By changing the criterion, one can increase or decrease the probability of the Type I error
at the cost of decreasing or increasing false negative or Type II error, which must also be defined relative to this criterion.

The language of hypothesis testing is built around a bias that false positives are worse than false negatives,
and consequently an emphasis on $p$-values rather than power.
In the context of early warning signals this is perilous --
it suggests that we would rather fail to predict a catastrophe than to sound a false alarm.
To avoid this linguistic bias and the introduction of an nuisance parameter on which to define statistical significance,
we propose the use of receiver operating characteristic (ROC) curves.

\subsection*{ROC Curves}
\begin{figure}
   \begin{center}
     \includegraphics[width=\linewidth]{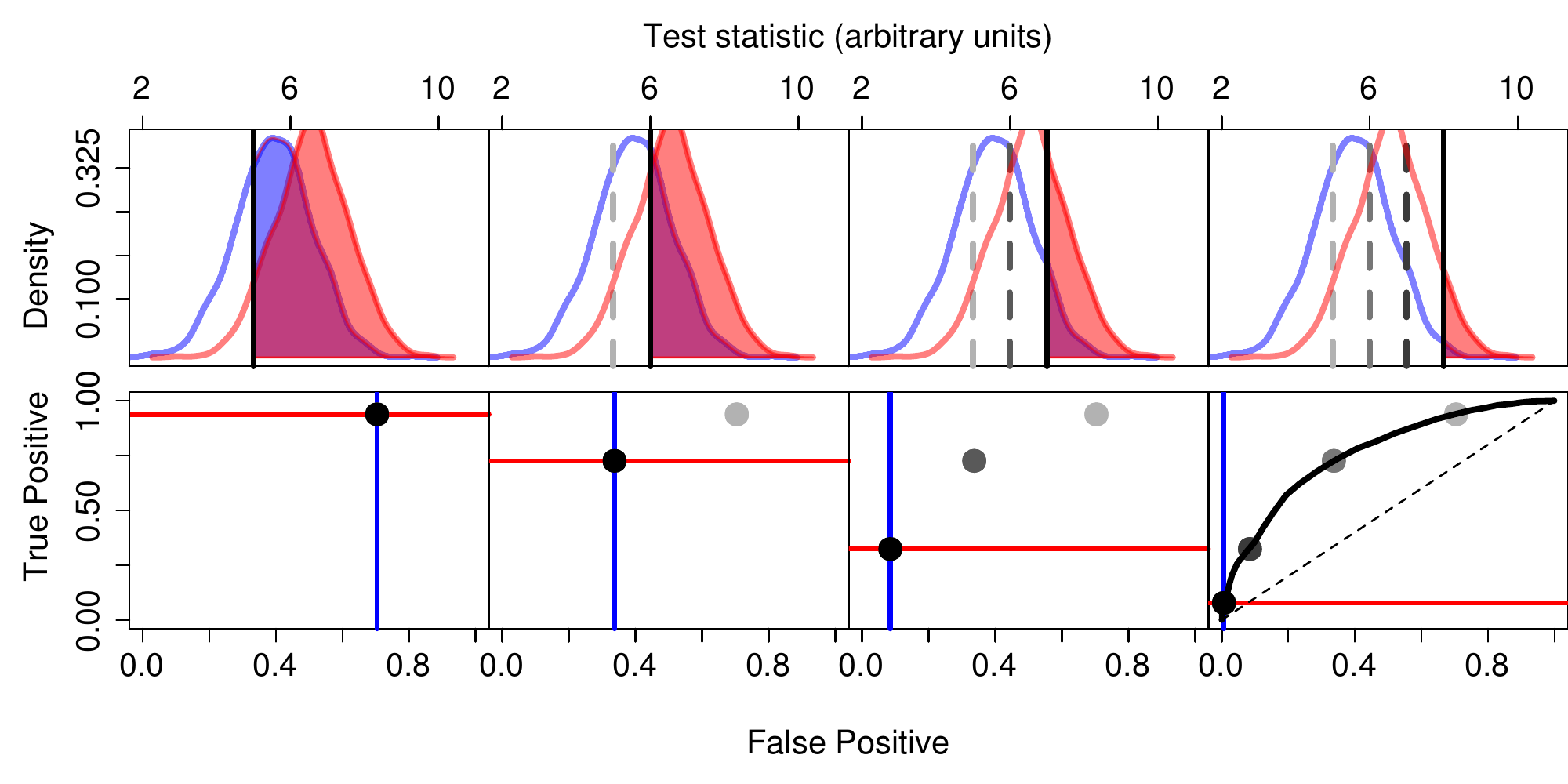}
     \caption{Top row: The distributions of a hypothetical warning indicator are shown under the case of a stable system (blue) and a system approaching a critical transition (red).  Bottom row: Points along the ROC curve are calculated for each possible threshold indicated in the top row.  The false positive rate is the integral of the distribution of the test statistic under the stable system right of the threshold (blue shaded area, corresponding to blue vertical line).  The true positive rate is the integral of the system approaching a transition left of the threshold (red shaded area, corresponds to the red line).  Successive columns show the threshold increasing, tracing out the ROC curve.}
     \label{fig:roc_intro}
  \end{center}
 \end{figure}

We illustrate the trade-off between false alarms and failed detection using
receiver-operating characteristic curves first developed in signal-processing literature~\citep{Green1989, Keller2009}⁠.
The curves represent the corresponding false alarm rate at any detection sensitivity (true positive rate), Fig~\ref{fig:roc_intro}.
The closer these distributions are to one-another, the more severe the trade-off.
If the distributions overlap exactly, the ROC curve has a constant slope of unity.
The ROC curve demonstrates this trade-off between accuracy and sensitivity.
Different early-warning indicators will vary in their sensitivity to detect differences
between stable systems and those approaching a critical transition,
making the ROC curves a natural way to compare their performance.
Since the shape of the curve will also depend on the duration and frequency of the time-series observations,
we can use these curves to illustrate by how much a given increase in sampling effort can decrease the rate of false alarms or failed detections.

\section{Example Results}
We illustrate this approach on simulated data as well as several natural time-series that have been previously analyzed for early warning signals.  All data and code for simulations and analysis are found in the accompanying R package, \verb|earlywarning|.

\subsection*{Data}
The simulation implements an individual, continuous-time stochastic birth-death process with rates given by the master equation~\citep{Gardiner2009},

\begin{align}
  \frac{dP(n,t)}{dt} &= b_{n-1} P(n-1,t) + d_{n+1}P(n+1,t) - (b_n+d_n) P(n,t)  \label{master} \\
    b_n &= \frac{e K n^2}{n^2 + h^2} \\
    d_n &= e n + a_t
\end{align}

where $P(n,t)$ is the probability of having $n$ individuals at time $t$, $b_n$ is the probability of a birth event occurring in a population of $n$ individuals, $d_n$ the probability of a death.  $e, K, h$ and $a_t$ are parameters.  
This corresponds to the well-studied ecosystem model of over-exploitation~\citep{Noy-Meir1975, May1977},
with stochasticity introduced directly through the demographic process.
We select this model since it is has discrete numbers of individuals, 
nonlinear processes, and the noise is driven by Poisson process of births and deaths instead of a Gaussian, 
and thus provides an illustration that our approach is robust to the violations of those
assumptions in model~\eqref{LSN}. 

\begin{figure}
   \begin{center}
     \includegraphics[width=.73\linewidth]{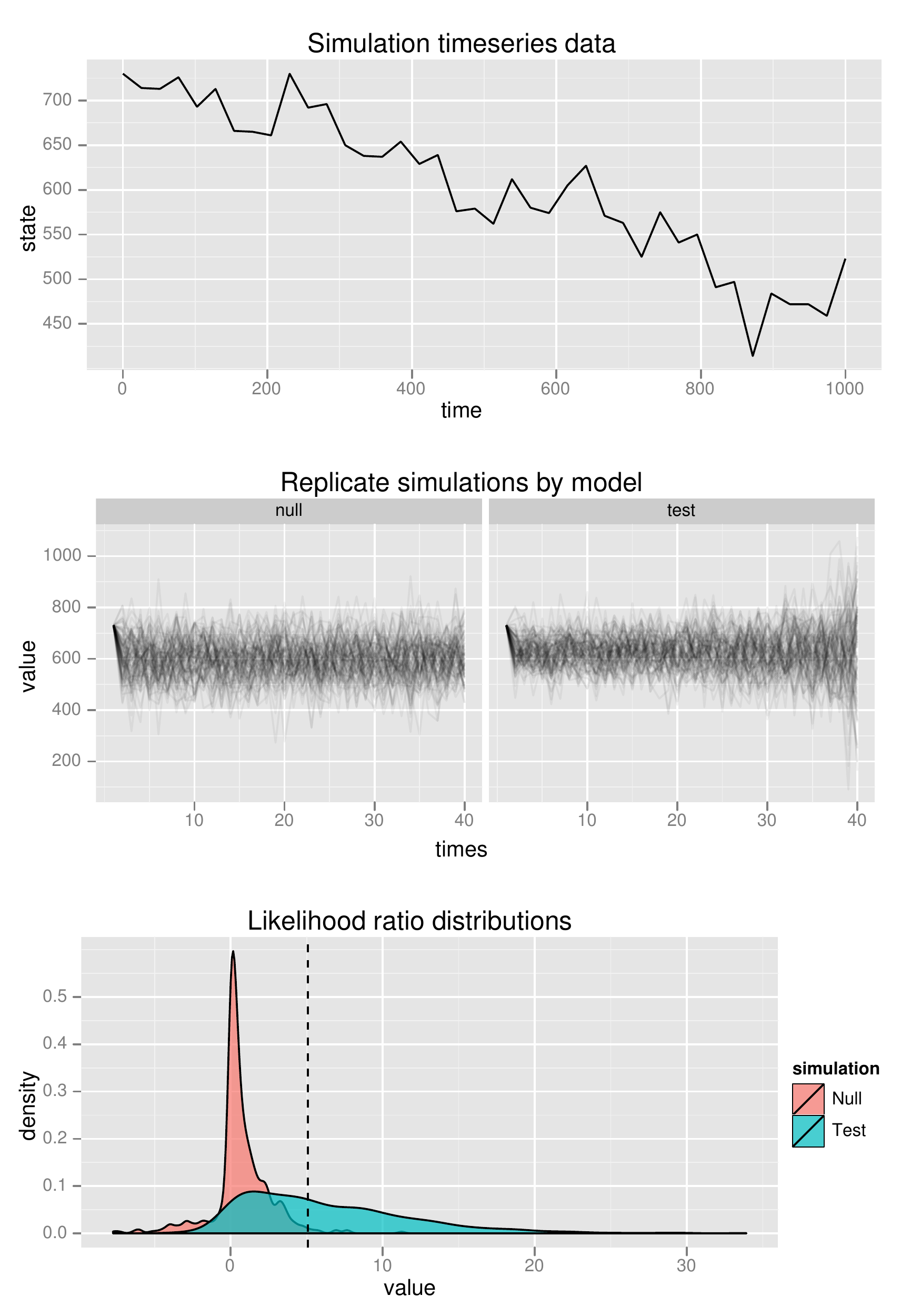}
     \caption{A model-based calculation of warning signals for the simulated data example.  Top panel: The original time series data on which model parameters for Equations~\eqref{OU}~and~\eqref{LSN} are estimated. Middle panel: replicate simulations under the maximum likelihood estimated (MLE) parameters of the null model, Equation~\eqref{OU} and test model, Equation~\eqref{LSN}.  Bottom panel: The distribution of deviances (differences in log likelihood, Equation~\eqref{delta}), when both null and test models are fit to each of the replicates from the null model, ``null,'' in red, and these differences when estimating for each of the replicates from the test model, in blue.  The overlap of distributions indicate replicates that will be difficult to tell apart.  The observed differences in the original data are indicated by the vertical line.}
     \label{fig:simulation}
  \end{center}
 \end{figure}

This model is forced through a bifurcation by gradually increasing the $a$ parameter,
which increases can be thought of as an increasing toxicity of the environment
(from $a_0 = 100$ increasing at constant rate of 0.09 units/unit time).
Other parameters are: $Xo = 730$, $e = 0.5$, $K = 1000$, $h = 200$.
We run this model over a time interval from 0 to 500
and sample at 40 evenly spaced time points, which were used for subsequent analysis.
This sampling frequency was chosen to be representative of reasonable sampling in biological time-series,
and provides enough points to detect a signal while not too many that errors can be avoided entirely.  
For the convenience of the inquisitive reader, we have also provided a simple function in the
associated R package where the user can vary the sampling scheme and parameter values and rerun this analysis.
This time series is shown in the top panel of Figure~\ref{fig:simulation}.


\begin{figure}
   \begin{center}
     \includegraphics[width=.85\linewidth]{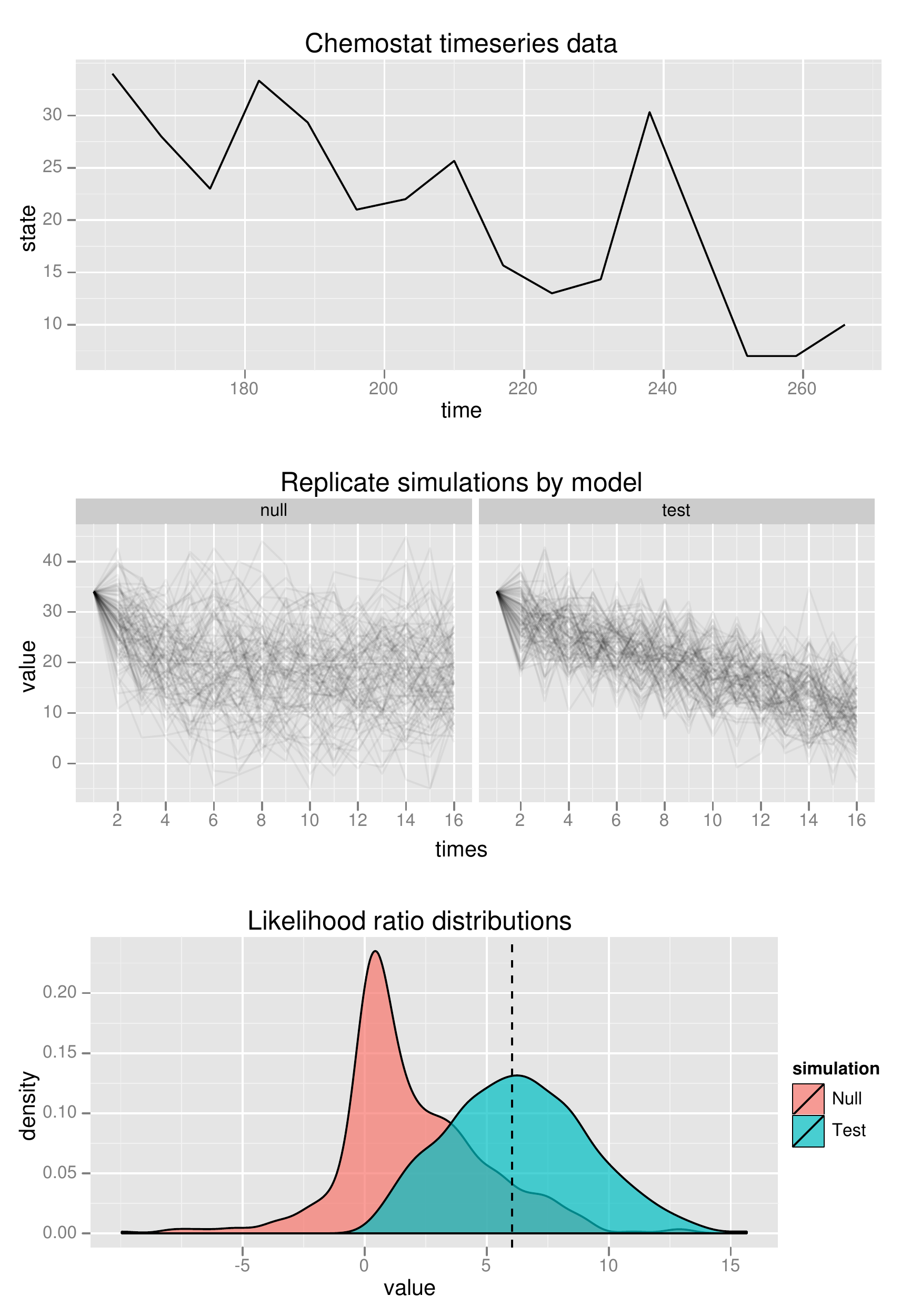}
     \caption{A model-based calculation of warning signals for the Daphnia data analyzed in ~\citet{Drake2010} (Chemostat H6).  
     Panels as in Figure~\ref{fig:simulation}. }
     \label{fig:chemostat}
  \end{center}
 \end{figure}

The first empirical data set comes from the population dynamics of
\emph{Daphnia} living in the chemostat ``H6'' in the experiments of Drake \& Griffen~\citep{Drake2010}.
This individual replicate was chosen as an example that showed
a pattern of increasing variance over the 16 data points where the system was being manipulated towards a crash.
This time series is shown in the top panel of Figure~\ref{fig:chemostat}.

\begin{figure}
   \begin{center}
     \includegraphics[width=.85\linewidth]{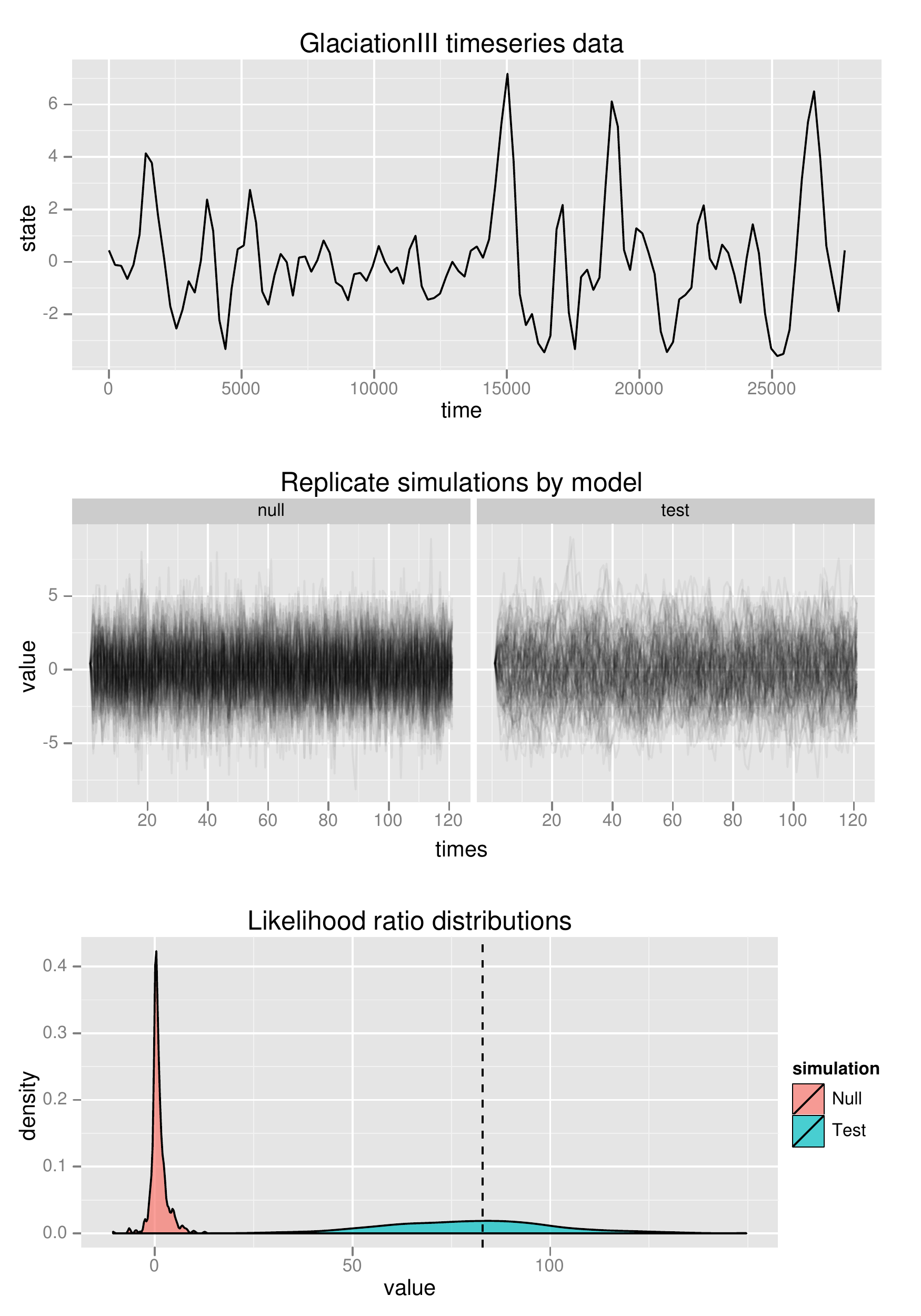}
     \caption{A model-based calculation of warning signals for the Glaciation data analyzed in ~\citet{Dakos2008} (Glaciation III).  Panels as in Figure~\ref{fig:simulation}.}
     \label{fig:glaciation}
  \end{center}
 \end{figure}

Our second empirical data set comes from the glaciation record seen
in deuterium levels in Antarctic ice cores~\citep{Petit1999},
as analyzed by~\citet{Dakos2008}.
The data are preprocessed by linear interpolation and de-trending by Gaussian kernel smoothing
to be as consistent as possible with the original analysis.
We focus on the third glaciation event, consisting of 121 sample points.
The match is not exact since~\citet{Dakos2008} estimates the de-trending window size manually,
but the estimated correlations in the first-order auto-regression coefficients are in close agreement with that analysis.
De-trending is intended to make the data consistent with the assumptions of the warning signal detection~\citep{Dakos2008},
which did not apply to the other data sets~\citep{Drake2010}.
This time series is shown in the top panel of Figure~\ref{fig:glaciation}.

\subsection*{Analysis}
The deviances $\delta$ observed are 5.1, 6.0, 83.9 for the simulation,
the chemostat data and the glaciation data, respectively.
Based on AIC score each is large enough to reject the null hypothesis of a stable model 
with its one extra parameter, but this does not give the full picture of the anticipated error rates.  
The size of these differences reflects not only the magnitude of the difference in fit
between the models but also the arbitrary units of the raw likelihoods,
which are smaller for larger data-sets.
Consequently the glaciation score reflects as much the greater length of its time series as it does anything else.

Our simulation approach can provide a better sense of the relative trade-off in error rates associated
with these estimates.  
As described above (Section~\ref{simbased}), we simulate 500 replicates under each model, 
shown in the middle panels of Figures~\ref{fig:simulation},~\ref{fig:chemostat} and~\ref{fig:glaciation},
and determine the distributions in likelihood ratio under each, shown in the lower panels.  
The observed deviance from the original data is also indicated (vertical line).

\begin{figure}
   \begin{center}
     \includegraphics[width=\linewidth]{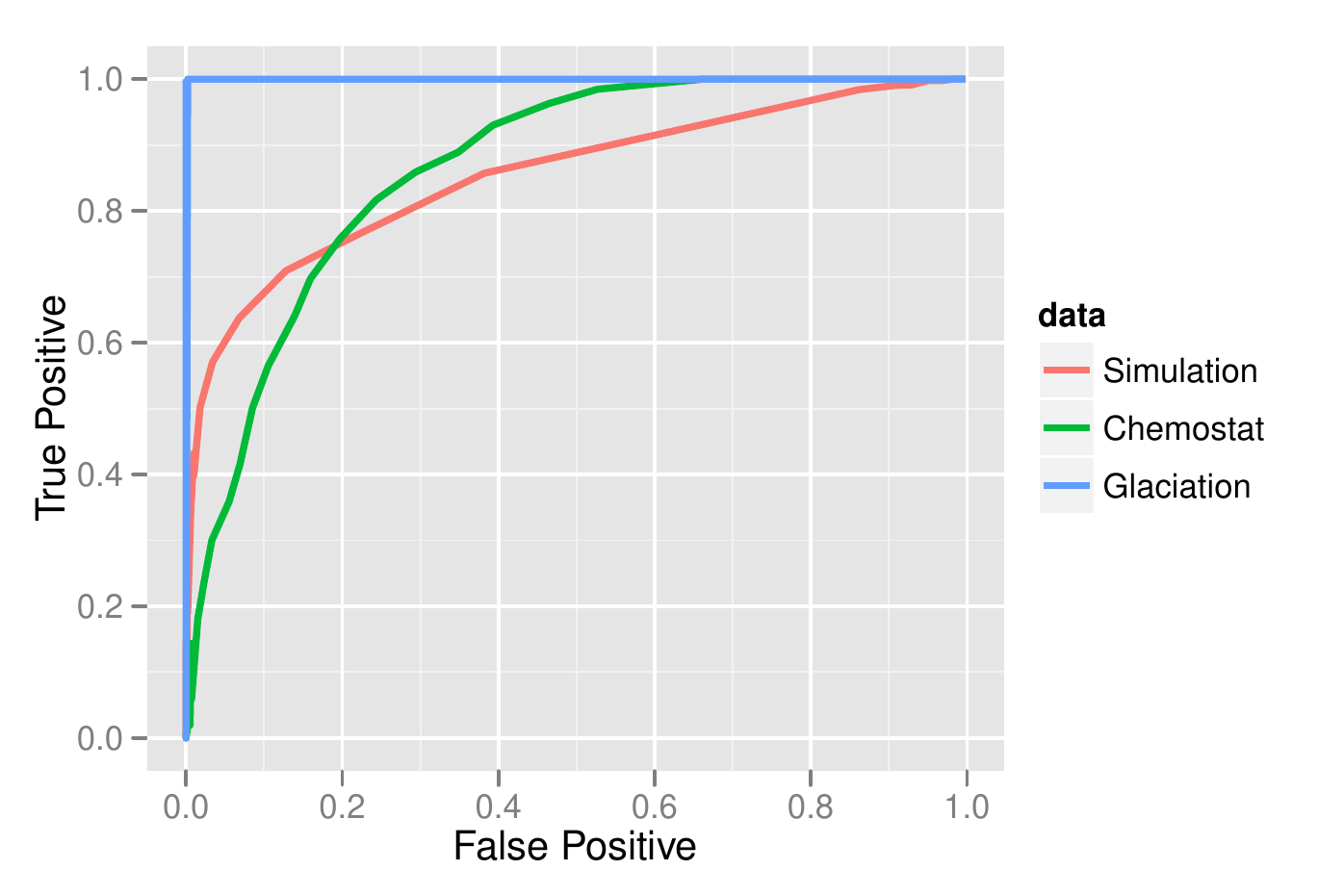}
     \caption{ROC curves for the Simulation, Chemostat, and Glaciation data, computed from the distributions shown in Figures~\ref{fig:simulation}, \ref{fig:chemostat} and \ref{fig:glaciation}, bottom panel.}
     \label{fig:rocdata}
  \end{center}
 \end{figure}

The ROC curves for each of these data sets are plotted in Figure~\ref{fig:rocdata}.  
While differences in the rate at which the system approaches a transition will also 
improve the ratio of true positives to false positives, 
here we see the best-sampled data set, Glaciation, with 121 points, 
also has the clearest signal with no observed errors in the 500 replicates of each type.   
Comparing the chemostat and simulation curves illustrate how the trade-off between
false positives and true positives can vary between data.  The chemostat signal, 
which estimates a relatively rapid rate of change but has less data, captures a higher
rate of true positives for a given rate of false positives than the simulation data
set with a weaker rate of change but more data, for false positive rates above 20\%.  
However, the simulated set with more data performs better if lower false-positive rates
are desired.


\section{Comparing the performance of summary statistics and model-based approaches}
Due to the variety of ways in which early warning signals based on summary statistics are implemented and evaluated
it is difficult to give a straight-forward comparison between them and the performance of this model-based approach.
However, by adopting one of one of the quantitative measures of a warning signal pattern,
such as Kendall's $\tau$~\citep{Dakos2008, Dakos2011, Dakos2009},
we are able to make a side-by-side comparison of the different summary statistics and the model based approach in the
context of false alarms and failed detections shown by the ROC curve.
Values of $\tau$ near unity indicate a strongly increasing trend in the warning indicator,
which is supposed to be indicative of an approaching transition.
Values near zero suggest a lack of a trend, as expected in stable systems.

\begin{figure}
   \begin{center}
     \includegraphics[width=\linewidth]{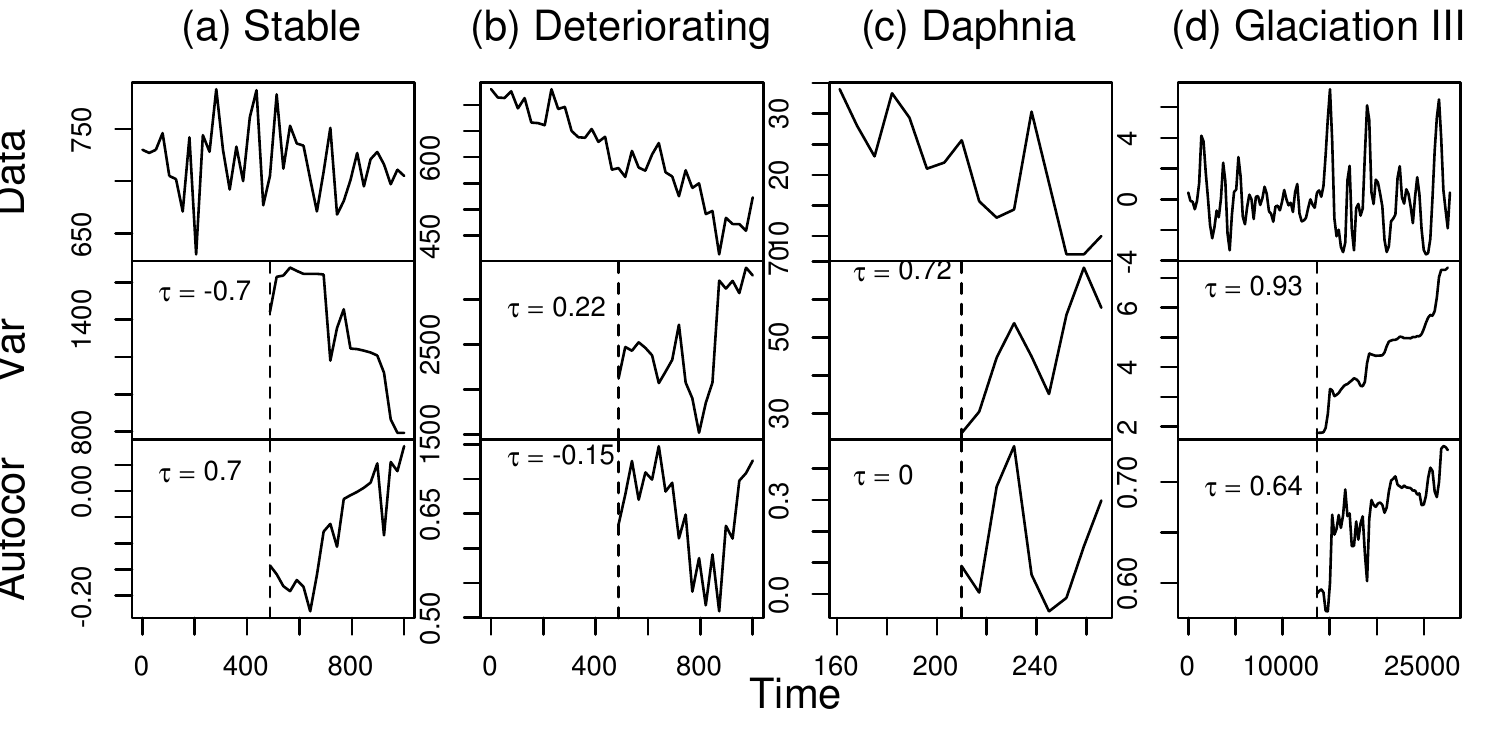}
     \caption{Early warning signals in simulated and empirical data sets.  
     The first two columns are simulated data from (a) a stable system (Stable), 
     and (b) the same system approaching a saddle-node bifurcation (Deteriorating).  
     Empirical examples are from (c) \emph{Daphnia magna} concentrations manipulated towards a critical transition (Daphnia), 
     and (d) deuterium concentrations previously cited as an early warning signal of a glaciation period (Glaciation). 
     Increases in summary statistics, computed over a moving window, 
     have often been used to indicate if a system is moving towards a critical transition.  
     The increase is measured by the correlation coefficient $\tau$.  
     Note that positive correlation does not guarantee the system is moving towards a transition, 
     as seen in the stable system, first column.}
     \label{fig:summary_stats}
  \end{center}
 \end{figure}

Figure~\ref{fig:summary_stats} shows the time series for each data set in columns
and the early warning indicators  of variance and autocorrelation
computed over a sliding window for each.
Kendall's correlation coefficient $\tau$ is calculated for each warning indicator and displayed on the graphs, inset.
For comparison, the left-most column includes data simulated under a stable system,
which nevertheless shows a chance increasing autocorrelation with a $\tau=0.7$
We can adapt the approach we have described above to determine how often such a strong increase would appear
by chance in a stable system as follows.

 \begin{figure}
   \begin{center}
     \includegraphics[width=\linewidth]{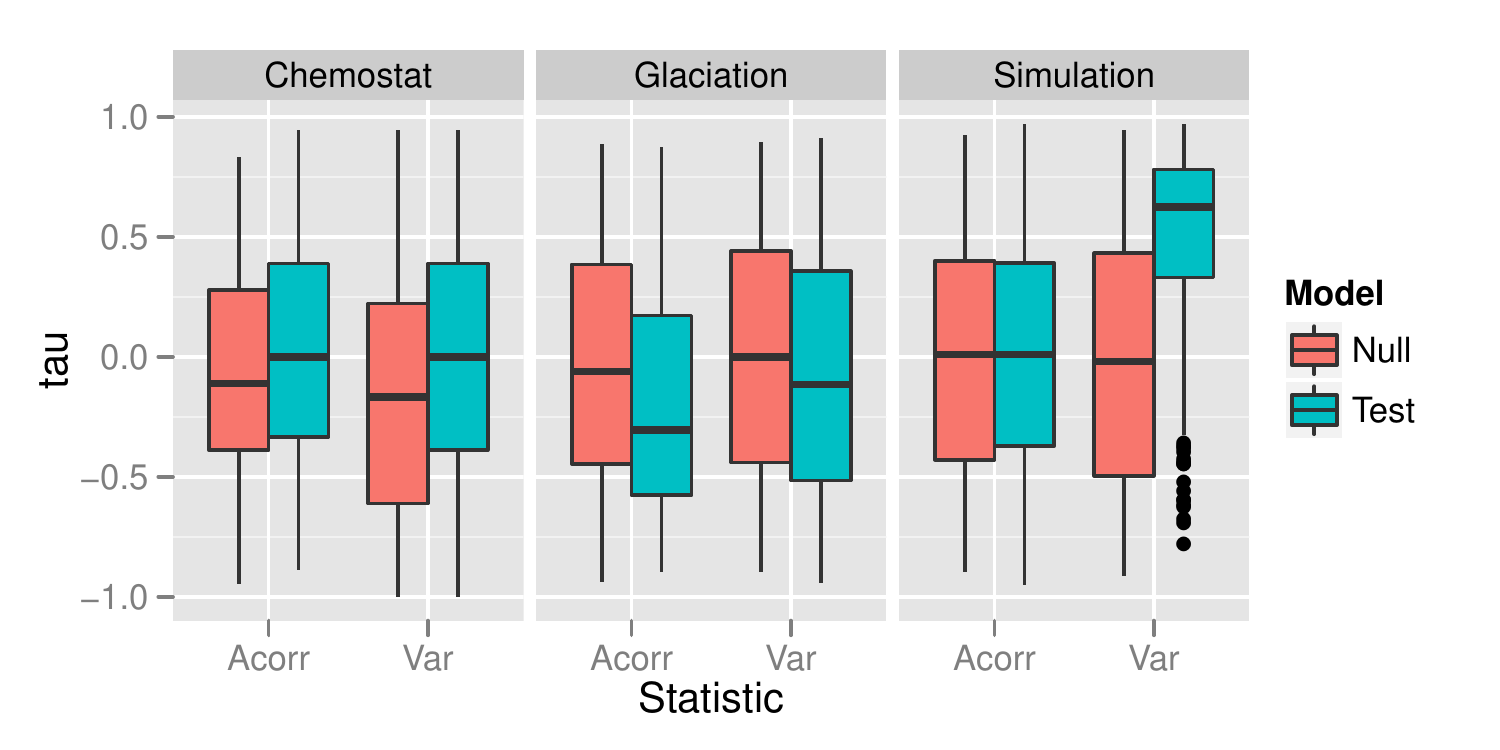}
     \caption{Box-plots of the distributions of Kendall's~$\tau$ observed for the summary statistic methods variance and autocorrelation, applied to three different data sets (from Figures~\ref{fig:simulation},~\ref{fig:chemostat}, \ref{fig:glaciation}).  The distributions show extensive overlap, suggesting that it will be difficult to distinguish early warning signals by the correlation coefficient in these summary statistics.}
     \label{fig:summary_box}
  \end{center}
 \end{figure}

By estimating the stable and critical transition models from the data,
and simulating 500 replicate data sets under each as in the analysis above,
we can then calculate the warning signals statistic over a sliding window
of size equal to one-half the length of the time series,
and compute the correlation coefficient $\tau$ measuring the degree to which the statistic shows an increasing trend.
This results in a distribution of $\tau$ values coming from a model of a stable system,
and a corresponding distribution of $\tau$ values coming from the model with an impending transition.
These distributions are shown in Figure~\ref{fig:summary_box}.
Contrary to the expectation that replicates of the null model (stable system, Equation~\eqref{OU}) would cluster around zero, 
while the test model, Equation~\eqref{LSN}, would cluster around larger positive $\tau$ values, 
the observed $\tau$ values on the replicates extend evenly across the range.  
This results in dramatic overlap and offers little ability to distinguish between the stable replicates 
and the replicates approaching a transition. 

The use of box plots in Figure~\ref{fig:summary_box} provide a convenient and familiar way 
to visualize the overlap between more than two distributions,
though they lack the resolution of the overlapping density distributions in 
Figures~\ref{fig:simulation},~\ref{fig:chemostat}, \ref{fig:glaciation}.  
The overlapping distributions are the natural representation from which to introduce the ROC curve, as in Figure~\ref{fig:roc_intro}.

 \begin{figure}
   \begin{center}
     \includegraphics[width=.8\linewidth]{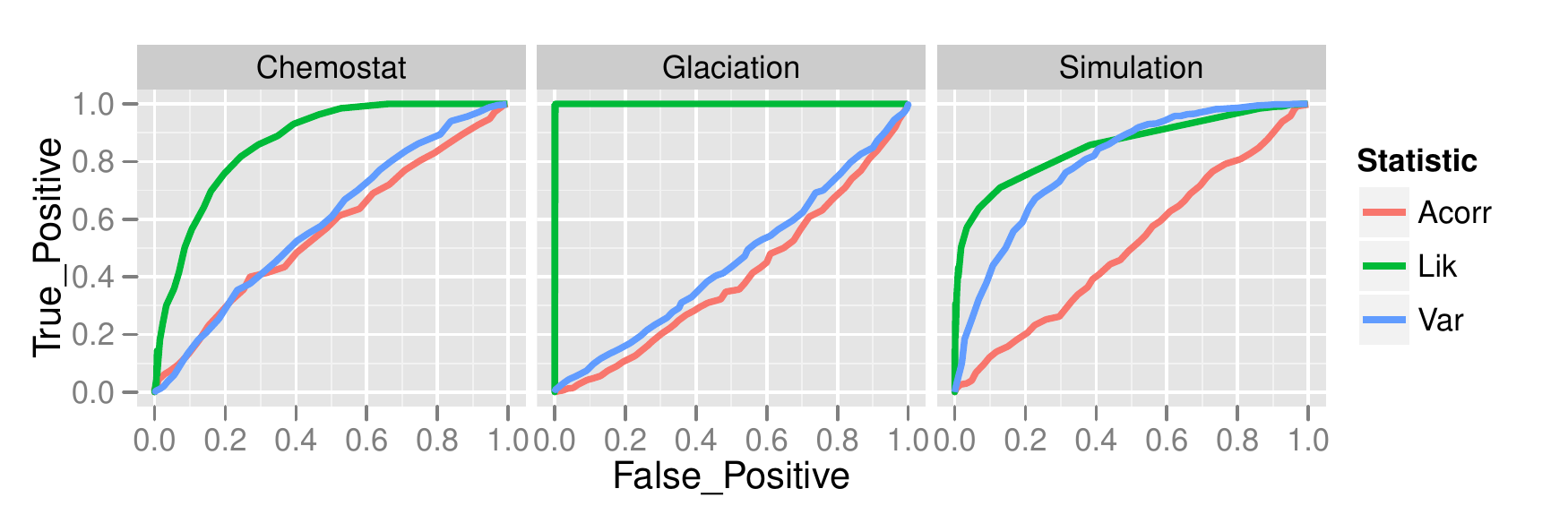}
     \caption{ROC curves compare the performance of the summary statistics variance and autocorrelation against the likelihood-based approach from Figure~\ref{fig:rocdata} on each of three example data sets (Figures~\ref{fig:simulation},~\ref{fig:chemostat}, \ref{fig:glaciation}).}
     \label{fig:summary_roc}
  \end{center}
 \end{figure}

 The ROC curves for these data (Fig.~\ref{fig:summary_roc}) show that the summary-statistic based indicators
frequently lack the sensitivity to distinguish reliably between observed patterns from a stable or unstable system.
The large correlations observed in the empirical examples (Fig.~\ref{fig:summary_stats}) are not uncommon in stable systems.
It is notable that in both empirical examples the summary statistics approach does little better than chance in distinguishing 
replicates that have been simulated from models~\eqref{LSN} and ~\eqref{OU}, 
despite the fact that these models correspond 
to the assumptions of the summary statistics approaches.  
On the simulated data, the variance based method approaches the 
true-positive rate of our likelihood method at higher levels of false positives, 
but performs worse when the desired level of false positives is low.  
The ROC curve helps us compare the performance of the different approaches at different tolerances.  
For instance, Table~\ref{tab:falsepos} shows the fraction of true crashes caught at a 5\% false positive rate.
We can instead set a desired True positive rate and read off the resulting number of false alarms, Table~\ref{tab:truepos}.  

\begin{table}
  \begin{center}
    \begin{tabular}{l|l|l}
                  &  Variance & Likelihood \\ 
      \hline
      Simulation  & 25 \%     & 61\% \\
      Chemostat   & 5.0\%     & 34\% \\
      Glaciation  & 5.4\%     & 100\% 
    \end{tabular}
    \caption{Fraction of \emph{crashes detected} when the desired false alarm rate is fixed to 5\%}
    \label{tab:falsepos}
  \end{center}
\end{table}

\begin{table}
  \begin{center}
    \begin{tabular}{l|l|l}
                  &  Variance & Likelihood \\ 
      \hline
      Simulation  & 49 \%     & 55\% \\
      Chemostat   & 81 \%     & 35\% \\
      Glaciation  & 93 \%     & 0\% 
    \end{tabular}
    \caption{Fraction of \emph{false alarms} when the desired detection rate is fixed to 90\%}
    \label{tab:truepos}
  \end{center}
\end{table}

\section{Discussion}

The challenge of determining early warning signs for impending possible regime shifts 
requires real attention to the underlying statistical issues and other assumptions.  
Doing this, does, however, open up new possibilities for asking what the goal of detection should be, 
and for clearly identifying underlying assumptions. 
We consider alternative approaches based either on summary statistics or a likelihood based model choice.  
By assuming the underlying model corresponds to a saddle-node bifurcation,
our analysis presents a ``best-case scenario'' for both  summary statistic and  likelihood-based approaches.
Other literature has already begun to address the additional challenges posed when the underlying
dynamics do not correspond to these models~\citep{Hastings2010}.
Our results illustrate that even in this best-case scenario,
reliable identification of warning signals from summary statistics can be difficult.

We have used three examples to illustrate the performance of this approach in data from 
simulation, a chemostat experiment, and paleo-atmospheric record; examples differing in 
sampling intensity and strength of signal of an approaching collapse.  
While the well-sampled geological data shows an unmistakable signal in this model-based approach,
the uncertainty in the smaller simulated and experimental data forces a trade-off between errors.  
 
As a way to clearly illustrate the choices involved in looking for warning signals while avoiding false alarms,
we introduce an approach based on receiver operator curves. 
These curves illustrate the extent to which an potential warning signal mitigates the trade-off 
between missed events and false alarms.  
The extent of the difficulty in finding reliable indicators of impending regime shifts
based on summary statistics becomes clear from the ROC curves of these statistics,
where a 5\% false positive rate often corresponds to only a 5\% true positive rate,
performing no better than the flip of a coin. 
By estimating the ROC curve for a given set of data,
we can better avoid applying warning signals in cases of inadequate power.
By taking advantage of the assumptions being made to write down a specific likelihood function,
we can develop approaches that get the most information from the data available.

In any application of early warning signals, it is essential to address the question of model adequacy.
Our approach formalizes the assumptions about the underlying process to match the assumptions of the other warning signals.
As the bifurcation results from the principle eigenvalue passing through zero,
the warning signal is expected in linear-order dynamics;
estimation of the nonlinear model is less powerful and less accurate.
The performance of this approach in the simulated data -- which is nonlinear in its dynamics
and driven with non-Gaussian noise introduced by the Poisson demographic events --
demonstrates the accuracy under violation of these assumptions.

The conclusion is not simply that likelihood approaches are more reliable,
but rather more broadly that warning signals should consider
the inherent trade-off between sensitivity and accuracy,
and must quantify how this trade-off depends on both the indicators used and the data available.
The approach developed here estimates the risk of both failed detection and false alarms;
concepts which are critical to prediction-based management.
Using the methods we have outlined when designing early warning strategies for natural systems
can ensure that data collection has adequate power to offer a reasonable chance of detection.

\section{Acknowledgments}
We would like to thank S. Schreiber, M. Holyoak, M. Baskett, A. Perkins, N. 
Ross, M. Holden, and two anonymous reviewers for comments on the manuscript. 
This research was supported by funding from NSF Grant EF 0742674 
and a Computational Sciences Graduate Fellowship from the Department of Energy grant DE-FG02-97ER25308. 
Data and code are available at https://github.com/cboettig/earlywarning.

\section{References}
\bibliography{boettiger}
 \end{document}